\documentstyle[editedvolume,numreferences,amsmath,amssymb,epsfig]{crckapb}

\newif\ifpreprint 
\preprinttrue

\numberwithin{equation}{section}
\renewcommand\theequation{\ifnum\arabic{section}=0
   \arabic{equation}\else\thesection.\arabic{equation}\fi}

\ifpreprint  \fi 

\newcommand\skipthis[1]{{}}

\newcommand\hepth[1]{{\sf hep-th/{#1}}}
\newcommand\phepth[1]{\ifpreprint{ {\sf[}\hepth{#1}{\sf]}}\fi}
\newcommand\grqc[1]{{\sf gr-qc/{#1}}}
\newcommand\pgrqc[1]{\ifpreprint{ {\sf[}\grqc{#1}}{\sf]}\fi}

\newcommand\slr{$SL(2,{\mathbb{R}})$}
\newcommand\half{\tfrac{1}{2}}

\newcommand\p{\partial}

\newcommand\anti[2]{{\bigl\{{#1},{#2}\bigr\}}}
\newcommand\com[2]{{\bigl[{#1},{#2}\bigr]}}
\DeclareMathOperator*{\Tr}{Tr}
\newcommand\ket[1]{{\lvert{#1}\rangle}}

\newcommand\lie[2]{{\pounds_{\!{#1}}{#2}}}
\newcommand\cs[1]{I^{{#1}}}
\newcommand\kf[1]{J^{{#1}}}

\newtheorem{ex}{Exercise}
\newtheorem{hardex}[ex]{$\ast$Exercise}

\begin{opening}

\title{Lectures on Superconformal Quantum Mechanics
 and Multi-Black Hole Moduli Spaces}%

\ifpreprint
\author{Ruth Britto-Pacumio$^1$}
\institute{}
\author{Jeremy Michelson$^2$}
\institute{}
\author{Andrew Strominger$^1$}
\institute{}
\author{Anastasia Volovich$^{1,3}$}
\institute{\\ $^1$Jefferson Physical Laboratories, Harvard University,
Cambridge, MA~~02138, USA \\ \\
$^2$New High Energy Theory Center, Rutgers University, 126~Frelinghuysen
Road, Piscataway, NJ~~08854, USA \\ \\
$^3$L.D.\ Landau Institute for
Theoretical Physics, Kosigina 2, 117334, Moscow, Russia}
\else
\author{Ruth Britto-Pacumio}
\institute{Jefferson Physical Laboratories \\ Harvard University \\
Cambridge, MA~~02138, USA}

\author{Jeremy Michelson}
\institute{New High Energy Theory Center \\ Rutgers University \\ 126
Frelinghuysen Road \\ Piscataway, NJ~~08854,  USA}

\author{Andrew Strominger}
\institute{Jefferson Physical Laboratories \\ Harvard University \\
Cambridge, MA~~02138,  USA}

\author{Anastasia Volovich}
\institute{Jefferson Physical Laboratories \\ Harvard University \\
Cambridge, MA~~02138,  USA ~~~
\\and
\\ L.D.\ Landau Institute for
Theoretical Physics,\\
 Kosigina 2, Moscow, Russia}
\fi

\end{opening}

\runningtitle{SCQM and Multi-Black Hole Moduli Spaces}
\ifpreprint \makeatletter \@runningauthorsetfalse \makeatother
\skipthis{\runningauthor{Ruth Britto-Pacumio et.\ al.}}
\runningauthor{Britto-Pacumio, Michelson, Strominger and Volovich}
\fi

\begin{document}

\ifpreprint 
\begin{abstract}
This contribution to the proceedings of the 1999 NATO ASI on
Quantum Geometry at Akureyri, Iceland, is based on notes of
lectures given by A. Strominger. Topics include $N$-particle
conformal quantum mechanics, extended superconformal quantum
mechanics and multi-black hole moduli spaces.
\end{abstract}
\fi

\ifpreprint \vfill HUTP-99/A060 \hfill hep-th/9911066 \hfill RUNHETC-99-41
\hfill
\newpage\fi

\ifpreprint \section{Introduction} \label{sec:intro} \else \section*{} \fi

The problem of unifying quantum mechanics and gravity is one of
the great unsolved problems in twentieth century physics. Progress
has been slowed by our inability to carry out relevant physical
experiments. Some progress has nevertheless been possible, largely
through the use of  gedanken experiments.

The quantum mechanical black hole has been a key ingredient of
these gedanken experiments, beginning with \cite{sh1,sh2}.  It
provides an arena in which quantum mechanics and gravity meet head
on. Such gedanken experiments have led to an astonishing depth and
variety of insights, not only about the black holes themselves,
but about string theory and quantum field theory in general.
Nevertheless many aspects of quantum black holes remain enigmatic,
and we expect they will continue to be a source of new insights.

Studies of quantum black holes have largely focused on the problem
of quantum fields or strings interacting (by scattering or
evaporation) with a single black hole. In these lectures we will
address a different, less studied,  type of gedanken experiment,
involving an arbitrary number $N$ of supersymmetric black holes.
Configurations of $N$ static black holes parametrize a moduli
space $\cal M_N$ \cite{fe,gr,ftr}. 
The low-lying quantum states of the system are
governed by quantum mechanics on $\cal M_N$. As we shall see the
problem of describing these states
has a number of interesting and puzzling features.
In particular $\cal M_N$ has noncompact, infinite-volume regions
corresponding to {\em near-coincident} black holes. These regions
lead to infrared divergences and presents a challenge for obtaining a
unitary description of multi-black hole scattering.

     The main goal of these lectures is to describe the recent
discovery of a superconformal structure 
\cite{ccj,rj,hag,nied}
in multi-black hole
quantum mechanics. While the appearance of scale invariance at low
energies follows simply from dimensional analysis, the appearance
of the full conformal invariance requires particular values of the
various couplings and is not {\em a priori} guaranteed. This structure
is relevant both to the infrared divergences and the scattering,
which however remain to be fully understood. We begin these
lectures by developing the subject of conformal and superconformal
quantum mechanics with $N$ particles. Section~\ref{dff} describes the
simplest example~\cite{dff} of single-particle conformally
invariant quantum mechanics. The infrared problems endemic to
conformal quantum mechanics as well as their generic cure are
discussed in this context. Section~\ref{cqm} contains a discussion of
conformally invariant $N$-particle quantum mechanics.
Superconformal quantum mechanics is described in section~\ref{sqm}. In
section~\ref{tp} the case of a test particle moving in a black hole
geometry is discussed (following \cite{bhaf}) as a warm-up to the
multi-black hole problem. The related issues of conformal
invariance, infrared divergences and choices of time coordinate
appear and are discussed in this simple context. In section~\ref{bhms} the
five dimensional multi-black hole moduli space as well as its
supersymmetric structure are described. It is shown that at low
energies the supersymmetries are doubled and the $D(2,1;0)$
superconformal group makes an appearance. We close with a
conjecture in section~\ref{conc} on the possible relation to an M-brane
description of the black hole and $AdS_2/$CFT$_1$ duality \cite{jm}.

Many of the results described herein appeared recently in
\cite{jmas1,jmas2}.

\section{A Simple Example of Conformal Quantum Mechanics} \label{dff}

Let us consider the following Hamiltonian~\cite{dff}:
\begin{equation} \label{dffh}
H = \frac{p^2}{2} + \frac{g}{2 x^2}.
\end{equation}
In order to have an energy spectrum that is bounded from below, it
turns out that  we need to take $g\geq -1/4$,
 but otherwise $g$ is an arbitrary
coupling constant, though, following~\cite{dff}, we will consider only $g>0$. 
Next introduce the operators
\begin{align} \label{dffdk}
D &= \tfrac{1}{2} (px + xp) &
K &= \tfrac{1}{2} x^2.
\end{align}
$D$ is known as the generator of dilations~--- it generates
rescalings $x{\rightarrow}{\lambda}x$ and
$p{\rightarrow}p/{\lambda}$~--- and $K$ is the generator of
special conformal transformations. These operators obey the
\slr\ algebra%
\begin{subequations} \label{sl2r}
\begin{align} \label{comdh}
\com{D}{H} &= 2iH, \\ \label{comdk}
\com{D}{K} &= -2iK, \\ \label{comhk}
\com{H}{K} &= -iD.
\end{align}
\end{subequations}
Since $D$ and $K$ do not commute with the Hamiltonian, they do not
generate symmetries in the usual sense of relating degenerate
states. Rather they can be used to to relate states with different
eigenvalues of $H$~\cite{ccj,rj,hag,nied,dff}.

\begin{ex} \label{ex:dffcont}
Show that for any quantum mechanics with operators obeying the \slr\
algebra~\eqref{sl2r}, that if $\ket{E}$ is a state of energy $E$, then
$e^{i \alpha D}
\ket{E}$ is a state of energy $e^{2 \alpha} E$.  Thus, if there is a state
of nonzero energy, then the spectrum is continuous.
\end{ex}

It follows from exercise~\ref{ex:dffcont} that the spectrum of the
Hamiltonian~\eqref{dffh} is continuous, and its eigenstates are
not normalizable. Hence it is awkward to describe the theory in
terms of $H$ eigenstates.

This problem is easily rectified. Consider the linear combinations
\begin{subequations} \label{lpml0}
\begin{gather} \label{lpm}
L_{\pm1} = \tfrac{1}{2} (a H - \frac{K}{a} \mp  i D) \\
\label{l0}
L_0 = \tfrac{1}{2} (a H + \frac{K}{a}),
\end{gather}
\end{subequations}
where $a$ is a parameter with dimensions of length-squared.
These obey the \slr\ algebra in the Virasoro form,
\begin{align} \label{pmsl2r}
\com{L_{1}}{L_{-1}} &= 2 L_0 &
\com{L_0}{L_{\pm1}} = \mp L_{\pm1}.
\end{align}
In the following, we choose our units such that $a=1$.

\begin{figure}[tb]
\begin{center}
\includegraphics[height=2in]{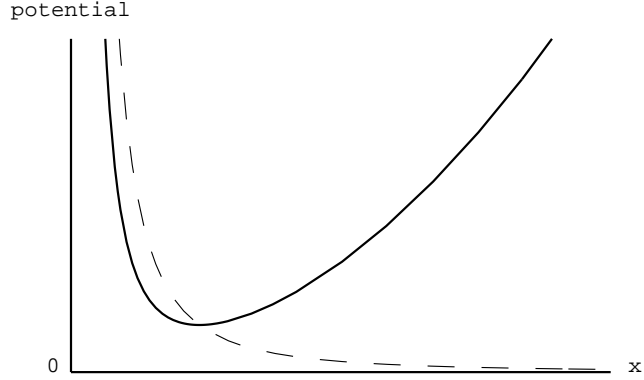}
\end{center}
\caption{A comparison between the potentials for $H$ and $L_0$.
The dashed line is the potential energy part of $H$ and the solid
line is that for $L_0$.  Note that the former has no minimum while
the latter is a well.} \label{fig:hvsl0pot}
\end{figure}

With the definitions~\eqref{l0}, \eqref{dffh} and~\eqref{dffdk}, we have
\begin{equation} \label{expl.l0}
L_0 = \frac{p^2}{4} + \frac{g}{4 x^2} + \frac{x^2}{4}.
\end{equation}
The potential energy part of this operator achieves its minimum and
asymptotes to $\infty$ (see
figure~\ref{fig:hvsl0pot}) and thus has a discrete spectrum with normalizable
eigenstates.

\begin{ex} \label{ex:cas.dff}
Show that
\begin{equation} \label{defl2}
L^2 = L_0(L_0-1) - L_{-1} L_1
\end{equation}
is the \slr\ Casimir operator.  Thus show that, of the eigenstates of
$L_0$, that with the smallest value of $L_0$ is annihilated by $L_1$.
Also show that the eigenvalues of $L_0$ form an infinite tower above the
``ground state'', in integer steps.
\end{ex}

\begin{hardex} \label{ex:gndl0}
Show that for the DFF model, the Casimir operator~\eqref{defl2} takes the
value
\begin{equation} \label{dffl2}
L^2 = \frac{g}{4} - \frac{3}{16}
\end{equation}
and thus that the ``ground state'' has $L_0 =
\frac{1}{2}(1\pm\sqrt{g+\frac{1}{4}})$.   (It turns out that the positive
root is that for which the state is normalizable.)
\end{hardex}

From exercise~\ref{ex:cas.dff}, we learn that the spectrum of
$L_0$ is well defined, and thus has normalizable eigenstates. This
motivated DFF to trade $H$ for $L_0$, and use $L_0$ to generate
the dynamics.  We then have a well defined theory; this also
justifies our use of the term ``ground state'' in
exercises~\ref{ex:cas.dff} and~\ref{ex:gndl0}.

At this point it is a free world and one has the right to describe
the theory in terms of $L_0$ rather than $H$ eigenstates. Later on
this issue will reappear in the context of black hole physics, and
the trade of $H$ for $L_0$ will take on a deeper significance.

\section{Conformally Invariant N-Particle Quantum Mechanics}
\label{cqm}

In this section, we find the conditions under which a general
$N$-particle quantum mechanics admits an \slr\ symmetry.
Specifically, we derive the conditions for the existence of
operators $D$ and $K$ obeying the algebra~\eqref{sl2r}.
$N$-particle quantum mechanics can be described as a sigma model
with an $N$-dimensional target space.
The general Hamiltonian is%
\footnote{In this and all subsequent expressions, the operator
ordering is as indicated.}
\begin{equation} \label{mnhm}
H=\half P_a^\dagger g^{ab} P_b+V(X),
\end{equation}
where $a,b=1,\dots,N$ and the metric $g$ is a function of $X$. The
canonical momentum $P_a$ obeys $\com{P_a}{X^b}=-i\delta_a^b$ and
$\com{P_a}{P_b}=0$, and is given by
\begin{equation} \label{defp}
P_a=g_{ab}\Dot X^b=-i\p_a.
\end{equation}

\begin{ex} \label{ex:pdag}
Given the norm $(f_1,f_2) = \int d^N X \sqrt{g} f_1^* f_2$, show that
\begin{equation} \label{pdagdefpdag}
P_a^\dagger = \frac{1}{\sqrt{g}} P_a \sqrt{g} = P_a - i \Gamma^b_{ba},
\end{equation}
where $\Gamma^c_{ab}$ is the Christoffel symbol built from the metric
$g_{ab}$, and the dagger denotes
Hermitian conjugation.  Thus, $H\Psi = (-\nabla^2 + V)\Psi$, for all
(scalar) functions $\Psi(X)$.
\end{ex}

We first determine the conditions under which the theory, defined
by equation~\eqref{mnhm}, admits a dilational symmetry of the
general form
\begin{equation} \label{yop}
\delta_D X^a=D^a(X).
\end{equation}
This symmetry is generated by an operator
\begin{equation} \label{hmt}
D = \half D^a P_a + \text{h.c.}
\end{equation}
which should obey equation~\eqref{comdh},
\begin{equation} \tag{\ref{comdh}}
\com{D}{H} = 2iH.
\end{equation}
From the definitions~\eqref{hmt}
and~\eqref{mnhm}, one finds
\begin{equation} \label{rgl}
\com{D}{H}=-\frac{i}{2} P_a^\dagger (\lie{D}{g^{ab}}) P_b
               -{i } \lie{D}{V}
               -\frac{i}{4} \nabla^2 \nabla_a D^a,
\end{equation}
where $\lie{D}{}$ is the usual Lie derivative obeying
\begin{equation} \label{oig}
\lie{D}{g_{ab}}=D^cg_{ab,c}+{D^c}{_{,a}}g_{cb}+{D^c}{_{,b}}g_{ac}.
\end{equation}
Comparing equations~\eqref{oig} and~\eqref{comdh} reveals that
a dilational
symmetry exists if and only if there exists a conformal Killing
vector $D$ obeying
\begin{subequations} \label{ford}
\begin{gather} \label{ldr}
\lie{D}{g_{ab}}=2g_{ab} \\ \intertext{and} \label{rrp}
\lie{D}{V} =-2 V.
\end{gather}
\end{subequations}
Note that equation~\eqref{ldr} implies the
vanishing of the last term of equation~\eqref{rgl}. A vector field $D$
obeying~\eqref{ldr} is known as a {\em homothetic} vector field, and the
action of $D$ is known as a {\em homothety} (pronounced h'MAWthitee).

Next we look for a special conformal symmetry
generated by an operator $K=K(X)$
obeying equations~\eqref{comdk} and~\eqref{comhk}:
\begin{align} \tag{\ref{comdk}}
\com{D}{K} &= -2iK, \\ \tag{\ref{comhk}}
\com{H}{K} &= -iD.
\end{align}
With equation~\eqref{hmt}, equation~\eqref{comdk} is equivalent to
\begin{equation} \label{kio}
\lie{D}{K}=2K,
\end{equation}
while equation~\eqref{comhk} can be written
\begin{equation} \label{ius}
D_a dX^a = dK.
\end{equation}
Hence the one-form $D$ is exact.
One can solve for $K$ as the norm of $D^a$,
\begin{equation} \label{glo}
K=\half g_{ab}D^aD^b,
\end{equation}
which is globally well defined.
We shall adopt the phrase ``closed homothety'' to refer to
a homothety whose associated one-form is closed and exact.

\begin{ex} \label{ex:closed=exact}
Show that conversely, given a vector field
$D^a$ obeying equation~\eqref{ldr} and $dD = 0$, that $D_a dX^a=dK$ where $K$ is
{\em defined} by equation~\eqref{glo}.  Thus, every ``closed homothety'' is
an ``exact homothety'', and there is no significance in our choice of phrase.
(We have chosen to use the phrase ``closed homothety'' in order to avoid
confusion with a discussion of, say, quantum corrections.)
\end{ex}

\begin{hardex} \label{ex:homo=noncompact}
Show that if a manifold admits a homothety (not necessarily closed), then
the manifold is noncompact.
\end{hardex}

We should emphasize that the existence of $D$ did {\em not}
guarantee the existence of $K$. It is not hard~\cite{jmas2} to find
examples of
quantum mechanics with a $D$ for which the corresponding unique
candidate for $K$ (by equation~\eqref{glo}) obeys neither
equations~\eqref{comhk} nor~\eqref{ius}. Indeed a generic
homothety is not closed.%
\footnote{One can find even four dimensional
theories that are dilationally, but not conformally, invariant by
including higher derivative terms; for a scalar field $\phi(x^\mu)$, the
Lagrangian \hbox{${\mathcal{L}} =
f(\frac{\partial^\mu \phi \partial_\mu \phi}{\phi^4}) \phi^4$} is
dilationally invariant for any function $f$, but it is conformally
invariant only for $f(y) = -\frac{1}{2}y - \frac{\lambda}{4!}$.\cite{rjlec}}

\section{Superconformal Quantum Mechanics} \label{sqm}
This section considers supersymmetric quantum mechanics with up to
four supersymmetries and superconformal extensions with up to
eight supersymmetries. In lower dimensions the Poincar\'{e} groups are
smaller and hence so are the supergroups. This implies a richer
class of supersymmetric structures for a given number of
supercharges. In particular, in one dimension we shall encounter
structures which cannot be obtained by reduction from higher
dimensions.

\subsection{A Brief Diversion on Supergroups} \label{sgp}

Roughly, a supergroup is a group of matrices that take the block form
\begin{equation} \label{defsg}
\left(\text{\begin{tabular}{c|c} $A$ & $F_1$ \\ \hline $F_2$ & $B$
\end{tabular}} \right),
\end{equation}
where $A,B$ are ordinary matrices, and $F_{1,2}$ are fermionic matrices.
We are interested in quantum mechanics with a
supersymmetry whose
supergroup includes \slr; that is, supergroups of the form
\begin{equation} \label{defoursg}
\left(\text{\begin{tabular}{c|c} \slr & fermionic \\ \hline fermionic &
R-symmetry \end{tabular}} \right).
\end{equation}
There are many such supergroups; these have been tabulated in
table~\ref{t:sg}.

\begin{table}[tb]
\begin{center}
\begin{tabular}{ccc} \hline
Superalgebra & Dimension (\#b,\#f) & $R$-symmetry \\ \hline
$Osp(1|2)$ & (3,2) & 1 \\ \hline
$SU(1,1|1)$ & (4,4) & $U(1)$ \\ \hline
$Osp(3|2)$ & (6,6) & $SU(2)$ \\ \hline
$SU(1,1|2)$ & (6,8) & $SU(2)$ \\
$D(2,1;\alpha), \alpha\neq-1,0,\infty$ & (9,8) & $SU(2)\times SU(2)$ \\ \hline
$Osp(5|2)$ & (13,10) & $SO(5)$ \\ \hline
$SU(1,1|3)$ & (12,12) & $SU(3) \times U(1)$ \\
$Osp(6|2)$ & (18,12) & $SO(6)$ \\ \hline
$G(3)$ & (17,14) & $G_2$ \\
$Osp(7|2)$ & (24,14) & $SO(7)$ \\ \hline
$Osp(4^*|4)$ & (16,16) & $SU(2)\times SO(5)$ \\
$SU(1,1|4)$ & (19,16) & $SU(4) \times U(1)$ \\
$F(4)$ & (24,16) & $SO(7)$ \\
$Osp(8|2)$ & (31,16) & $SO(8)$ \\ \hline \hline
$Osp(4^*|2n), n>2$ & ($2n^2+n+6$, $8n$) & $SU(2) \times Sp(2n)$ \\ \hline
$SU(1,1|n),n>4$ & ($n^2+3$, $4n$) & $SU(n) \times U(1)$ \\ \hline
$Osp(n|2),n>8$ & ($\frac{1}{2}n^2-\frac{1}{2}n+3$, $2n$) & $SO(n)$ \\
\hline
\end{tabular}
\caption{The simple supergroups that contain an \slr\ subgroup
(see also~\protect\cite{kt}). The table is divided into those
which have eight or fewer (ordinary) supersymmetries (including
the exceptional supergroups) and those which have more than eight
(ordinary) supersymmetries (for which there are no exceptional
supergroups). The algebra of $Osp(4^*|2m)$ has bosonic part
$SO^*(4)\times Usp(2m)$, where $SO^*(4)\protect\cong
SL(2,{\mathbb{R}})\times SU(2)$ is a noncompact form of the
$SO(4)$ algebra. \label{t:sg}}
\end{center}
\end{table}

One simple series of supergroups is the $Osp(m|n)$ series; the elements
of $Osp(m|n)$ have the form
\begin{equation} \label{ospmn}
\left(\text{\begin{tabular}{c|c} $Sp(n)$ & fermionic \\ \hline fermionic & $SO(m)$
\end{tabular}} \right).
\end{equation}
Since $Sp(2) \cong$ \slr\ 
\footnote{The notation is such that only $Sp(2n)$ exist.} we are
interested in $Osp(m|2)$.  The simplest of these is $Osp(1|2)$,
which is a subgroup of the others.  We will describe the models
with this symmetry group, for the supermultiplet defined in
section~\ref{smult},
in section~\ref{osp12}.  We will skip $Osp(2|2) \cong SU(1,1|1)$~
\footnote{The supergroup $U(m,n|p)$ is generated by matrices of the
form~\eqref{defsg}, with \hbox{$A\in U(m,n)$} and \hbox{$B\in U(p)$}.
The subalgebra in which the matrices also obey $\Tr{A}=\Tr{B}$ generates
$SU(m,n|p)$.  However, with this definition, $SU(m,n|p=m+n)$ is not even
semisimple, for the identity matrix obeys $\Tr{A}=\Tr{B}$ and generates a
$U(1)$
factor.  The quotient
\hbox{$PSU(m,n|m+n) \cong SU(m,n|m+n)/U(1)$} {\em is} simple, and is often
denoted just
$SU(m,n|m+n)$, as we have done for $SU(1,1|2)$.\label{ft:umnp}}%
~--- these models were described in~\cite{jmas1}~--- and go directly
to $Osp(4|2)$
.  In fact, it will turn out that, for the supermultiplet we
consider, we will naturally obtain $D(2,1;\alpha)$ as the symmetry
group, where $\alpha$ is a parameter that depends on the target
space geometry.  $Osp(4|2)$ is the special case of $\alpha=-2$,
and appears, for example, when the target space is flat. The black
hole system described in section~\ref{bhms}
will turn out to have $D(2,1;0)$ superconformal symmetry.%
\footnote{$D(2,1;0)$ (and $D(2,1;\infty)$) is omitted from
Table~\ref{t:sg} because it is the semidirect product $SU(1,1|2)\rtimes
SU(2)$ and is therefore not simple.}
We will
explain this statement, and describe $D(2,1;\alpha)$ in more detail, in
section~\ref{d21a}.  First, we should describe the supermultiplet under
consideration.

\subsection{Quantum Mechanical Supermultiplets}
\label{smult}

There are many supermultiplets that one can construct in
one dimension.  In particular, unlike in higher dimensions, the
smaller  supersymmetry group does not require a matching of the
numbers of bosonic and fermionic fields. Much of the
literature~--- see, {\em e.g.}~\cite{wi,ew2,ew3,fr,salom,jg}~---
concerns the so-called type A multiplet, with a real boson and
complex fermion $(X^a,\psi^a)$, which can be obtained by
dimensional reduction of the \hbox{$1+1$ dimensional}
\hbox{${\mathcal{N}}=(1,1)$} multiplet.

This is not the multiplet we will consider here. For the black
hole \hbox{physics} that we will eventually consider, each black
hole will have four bosonic (translational) degrees of freedom, as
well as four fermionic degrees of freedom from the breaking of
one half of the minimal (8 supercharge) supersymmetry in five
dimensions.%
\footnote{Recently, four-dimensional black holes have
been described using a multiplet with 3 bosons and 4 fermions
\cite{alexmark}.}
Thus, we will consider the type B multiplet,
consisting of a real boson and a {\em real} fermion
$(X^a,\lambda^a=\lambda^{a\dagger})$. The supersymmetry
transformation, parametrized by a real Grassman parameter
$\epsilon$, is given by
\begin{align} \label{ds1}
\delta_\epsilon X^a &= -i \epsilon \lambda^a &
\delta_\epsilon \lambda^a &= \epsilon \Dot{X}^a,
\end{align}
where the overdot denotes a time derivative.

\begin{ex} \label{ex:ds1}
In an ${\cal N}=1$ superspace formalism, the type B multiplet is
given by a real supermultiplet
$X^a(t,\theta)=X^{a\dagger}(t,\theta)$, where $\theta$ is the
(real) fermionic coordinate, and we use the standard convention in
which the lowest component of the superfield is
notationally almost indistinguishable from the superfield itself. In
components, we write
\begin{equation} \label{defsx}
X^a(t,\theta) = X^a(t) - i \theta \lambda^a(t).
\end{equation}
The generator of supersymmetry transformations, $Q$ (which obeys
$Q^2=H=i\frac{d}{dt}$) is given by
\begin{equation} \label{defsq}
Q = \frac{\p}{\p\theta} + i \theta \frac{d}{dt}.
\end{equation}
Show that
\begin{equation} \label{dsx}
\delta_\epsilon X^a = \com{\epsilon Q}{X^a},
\end{equation}
as expected.  Note also that $Q=Q^\dagger$, and thus both sides of
equation~\eqref{dsx} are, indeed, real.
For completeness, we define the superderivative
\begin{equation} \label{defd}
D = \frac{\p}{\p\theta} - i \theta \frac{d}{dt},
\end{equation}
which obeys $D^2=-i\frac{d}{dt}$ and $\anti{D}{Q}=0$.
\end{ex}

As we have
already mentioned, there are many more multiplets than just the type B one;
see {\em e.g.}~\cite{cp,alexmark}.

\subsection{$O\lowercase{sp}(1|2)$--Invariant Quantum Mechanics} \label{osp12}

We now proceed to the simplest superconformal quantum mechanics
for the Type B supermultiplet defined in the previous subsection.
As in section~\ref{cqm}, we use a Hamiltonian formalism.

In general, the supercharge takes the form
\begin{equation} \label{defq}
Q=\lambda^a\Pi_a - \frac{i}{3} c_{abc} \lambda^a \lambda^b \lambda^c,
\end{equation}
where we define
\begin{equation} \label{defpi}
\Pi_a \equiv
P_a-\frac{i}{2}\omega_{abc}\lambda^b\lambda^c+\frac{i}{2}c_{abc}\lambda^b
\lambda^c \equiv P_a - \frac{i}{2} \Omega^+_{abc} \lambda^b \lambda^c,
\end{equation}
where $\omega_{abc}$ is the spin connection with the last two indices
contracted with the vielbein, and $c_{abc}$ is a (so-far) general 3-form.
The Hamiltonian is then given by
\begin{equation} \label{hfromq}
H = \half \anti{Q}{Q}.
\end{equation}
We remark that the bosonic part of this Hamiltonian is the special case of
equation~\eqref{mnhm} with $V=0$.

\begin{ex} \label{sltol}
Show that the most general, renormalizable superspace
action~\cite{cp}
\begin{equation} \label{sl}
S = i \int dt d\theta \left\{\half g_{ab} DX^a \Dot{X}^b + \frac{i}{6}
c_{abc} DX^a DX^b DX^c \right\},
\end{equation}
is given in terms of the component fields by
\begin{equation} \label{l}
S = \int dt \left\{ \half g_{ab} \Dot{X}^a \Dot{X}^b
+ \frac{i}{2} \lambda^a \left(g_{ab}\frac{D\lambda^b}{dt} - \Dot{X}^c
c_{abc} \lambda^b \right)
- \frac{1}{6} \p_d c_{abc} \lambda^d \lambda^a \lambda^b \lambda^c\right\},
\end{equation}
where
\begin{equation} \label{covd}
\frac{D\lambda^a}{dt} \equiv \dot{\lambda}^a + \dot{X}^b
\Gamma^a_{bc} \lambda^c,
\end{equation}
is the covariant time-derivative.
(Note that $g_{ab}=g_{(ab)}$ and $c_{abc}=c_{[abc]}$ are arbitrary (though
$g_{ab}$ should be positive definite for positivity of the kinetic energy)
functions of the superfield; {\em e.g.}
$g_{ab} = g_{ab}(X(t,\theta))$.)
In terms of $\lambda^\alpha \equiv \lambda^a e^\alpha_a$, show that
the action~\eqref{l} is
\begin{multline} \label{usel}
S = \int dt \left\{ \half g_{ab} \Dot{X}^a \Dot{X}^b
+ \frac{i}{2} \delta_{\alpha \beta} \lambda^\alpha \frac{D\lambda^\beta}{dt} -
\frac{i}{2} \dot{X}^c c_{c \alpha \beta  } \lambda^\alpha  \lambda^\beta
\right. \\ \left.
- \frac{1}{6} e^d_\delta \p_d c_{abc}e^a_\alpha e^b_\beta e^c_\gamma
\lambda^\delta \lambda^\alpha \lambda^\beta \lambda^\gamma\right \},
\end{multline}
where
\begin{equation} \label{usecovd}
\frac{D\lambda^\alpha}{dt} = \dot{\lambda}^\alpha +
\dot{X}^a \omega_a{^\alpha}{_\beta} \lambda^\beta.
\end{equation}
$\ast$Finally, show that equation~\eqref{defq} follows from
equation~\eqref{ds1}
(or~\eqref{dsx}).
\end{ex}

We note that, from equation~\eqref{usel}, the fermions $\lambda^\alpha$
obey the
canonical anticommutation relation
\begin{equation} \label{ll}
\anti{\lambda^\alpha}{\lambda^\beta} = \delta^{\alpha\beta},
\end{equation}
and commute with $X^a$ and $P_a$.%
\footnote{Note that this implies that (generically) $\lambda^a$ does {\em
not} commute with $P_b$, but rather,
$\com{P_a}{\lambda^b}=-i(\omega_a{^b}{_c} - \Gamma^b_{ac})\lambda^c$.}
It follows from equation~\eqref{ll}, that the fermions can be represented
on the Hilbert space by $\lambda^\alpha=\gamma^\alpha/\sqrt{2}$, where
$\gamma^\alpha$ are the $SO(n)$ $\gamma$-matrices ($n$ is the
dimension of the target space), and that the wavefunction
is an $SO(n)$ spinor.
Thus $\Pi_a$ is just the covariant derivative
(with torsion $c$~--- see appendix~\ref{torsion} for a brief summary of
calculus
with torsion) on the Hilbert space.%
\footnote{It also follows~\cite{ag,fw} that, for these theories, the Witten
index, $\Tr(-1)^F$~\cite{wi,ew2}, is
equivalent to the Atiyah-Singer index.}

So far we have only discussed ${\mathcal{N}}=1$ super{\em symmetric}
quantum mechanics,
whereas we would like to discuss super{\em conformal} quantum mechanics.
We have already shown in section~\ref{cqm} that in order to have conformal
quantum mechanics, the metric $g_{ab}$ must admit a closed homothety
$D^a$, out of which were built the operators $D$ and $K$.  The
supersymmetric extensions of the expressions~\eqref{hmt} and~\eqref{glo}
for $D$ and $K$~--- that is
including fermions~--- are given by replacing the $P_a$ in
equation~\eqref{hmt} (which is a covariant derivative on the scalar
wavefunction of the bosonic theory) with the covariant derivative $\Pi_a$:%
\footnote{But the reader should not extrapolate too far, for $H\neq \half
\Pi_a^\dagger g^{ab} \Pi_b$.}
\begin{gather} \label{sd}
D = \half D^a \Pi_a + \text{h.c.} \\
\intertext{and} \label{sk}
K = \half D^a D_a.
\end{gather}
However, it turns out that closure of the superalgebra places two
constraints on
the torsion $c_{abc}$,
\begin{subequations} \label{cfor1}
\begin{gather} \label{nc}
D^a c_{abc} = 0 \\ \intertext{and}
\label{liec}
\lie{D}{c_{abc}} = 2 c_{abc}.
\end{gather}
\end{subequations}

The final operator that appears in an $Osp(1|2)$-invariant
theory is
\begin{equation} \label{defs}
S = i\com{Q}{K} = \lambda^a D_a.
\end{equation}

\begin{hardex} \label{vops12}
Verify that (with equations~\eqref{cfor1}) the operators $H$, $D$, $K$, $Q$
and $S$, defined by equations~\eqref{hfromq}, \eqref{sd}, \eqref{sk},
\eqref{defq} and~\eqref{defs} satisfy the $Osp(1|2)$ algebra
\begin{equation} \label{osp12alg}
\begin{aligned}
\com{H}{K} &= iD,\qquad & \com{H}{D} &= -2iH,\qquad& \com{K}{D} &= 2iK, \\
\anti{Q}{Q} &= 2H,& \com{Q}{D} &= -iQ,& \com{Q}{K} &= -iS, \\
\anti{S}{S} &= 2K,& \com{S}{D} &= iS,& \com{S}{H} &= iQ, \\
\anti{S}{Q} &= D,& \com{Q}{H} &= 0,& \com{S}{K} &= 0.
\end{aligned}
\end{equation}
\end{hardex}

\subsection{$D(2,1;\alpha)$--Invariant Quantum Mechanics} \label{d21a}

The $D(2,1;\alpha)$ algebra is an ${\mathcal{N}}=4$ (actually
${\mathcal{N}}=4$B, since we use the type~B supermultiplet) superconformal
algebra,
and thus contains four supercharges $Q^m$, $m=1,\dots,4$, and their
superconformal partners $S^m$.  Of course, for fixed $m$, $Q^m,S^m,H,K,D$
should satisfy the $OSp(1|2)$ algebra~\eqref{osp12alg}.
In addition, as is evident from
table~\ref{t:sg}, there are two (commuting) sets of $SU(2)$ R-symmetry
generators $R_\pm^r$, $r=1,2,3$, under which the supercharges $Q^m$ and
$S^m$ transform as (2,2).  There are no other generators, and the complete
set of (anti)commutation
relations, which define the algebra, are~\cite{stv}
\begin{equation} \label{d21alg} \raisetag{2.5\baselineskip}
\begin{split}
\begin{aligned}
\com{H}{K} &= iD, & \com{H}{D} &= -2iH,& \com{K}{D} &= 2iK, \\
\anti{Q^m}{Q^n} &= 2H\delta^{mn},& \com{Q^m}{D} &= -iQ^m,& \com{Q^m}{K} &=
-iS^m, \\
\anti{S^m}{S^n} &= 2K\delta^{mn},& \com{S^m}{D} &= iS^m,& \com{S^m}{H} &=
iQ^m, \\
\com{R^r_\pm}{Q^m} &= it^{\pm r}_{mn} Q^n,\quad&
  \com{R^r_\pm}{S^m} &= it^{\pm r}_{mn} S^n,\quad&
  \com{R^r_\pm}{R^s_{\pm'}} &= i \delta_{\pm \pm'} \epsilon^{rst} R_\pm^t \\
\com{R^r_\pm}{H} &= 0,& \com{R^r_\pm}{D} &= 0,& \com{R^r_\pm}{K} &= 0, \\
\com{Q^m}{H} &= 0,& \com{S^m}{K} &= 0,
\end{aligned} \\
{\anti{S^m}{Q^n} = D\delta^{mn} -
\frac{4\alpha}{1+\alpha} t^{+r}_{mn} R^r_+ - \frac{4}{1+\alpha}
t^{-r}_{mn} R_-^r, \qquad\qquad}
\end{split}
\end{equation}
where
\begin{equation} \label{tpm}
t^{\pm r}_{mn} \equiv \mp \delta^r_{[m} \delta^4_{n]} + \half
\epsilon_{rmn}.
\end{equation}
Clearly the $D(2,1;\alpha)$ algebra is not defined for $\alpha=-1$; for
$\alpha=0$ ($\infty$), $R^r_+$ ($R^r_-$) does not appear on the right-hand
side of the commutation relations~\eqref{d21alg}, and thus the group is the
semidirect product of $SU(1,1|2)$~
\footnote{See footnote~\ref{ft:umnp} (page~\pageref{ft:umnp}) for the
definition of $SU(m,n|p)$.}
(the unique group in table~\ref{t:sg}
with the correct
number of generators and bosonic subalgebra) and $SU(2)$.

Before we discuss the conditions under which the action~\eqref{usel} admits
a $D(2,1;\alpha)$ superconformal symmetry, we should first discuss the
conditions for ${\mathcal{N}}=4$B supersymmetry.

\subsubsection{${\mathcal{N}}=4$B Supersymmetric Quantum Mechanics.}
The conditions on the geometry for an ${\mathcal{N}}=4$B theory have been
given in~\cite{cp,gps}.  We will repeat them in the simplified form given
in~\cite{jmas1}.  The object is to find supercharges $Q^m$, such that
\begin{equation} \label{qmtoh}
\anti{Q^m}{Q^n} = 2 \delta^{mn} H.
\end{equation}
We take $Q^4$ to be the $Q$ of equation~\eqref{defq}~--- {\em i.e.} the
Noether
charge associated with the symmetry generated by equation~\eqref{ds1}.  We
now look for three more symmetry transformations such that
\begin{equation} \label{dsalg}
\com{\delta^{(m)}_{\epsilon}}{\delta^{(n)}_{\eta}} = -2 i \delta_{mn}
{\eta}{\epsilon} \frac{d}{dt},
\end{equation}
where $\delta^{(m)}_\epsilon$ is the $m^{\text{\underline{th}}}$
supersymmetry transformation, generated by the Grassmann variable
$\epsilon$.  It is standard (see {\em e.g.}~\cite{ghr,cp,gps}) to
give these transformations according to the following rather tedious exercise.

\begin{ex} \label{extsusy}
Define
\begin{equation} \label{dsrest}
\delta^{(r)}_\epsilon X^a(t,\theta) = \epsilon \cs{r}_b{^a} D X^b,
\end{equation}
where $\cs{r}_b{^a}(X(t,\theta))$ is some tensor-valued function on
superspace.
Then, show that the supersymmetry algebra~\eqref{dsalg} is obeyed iff
$\cs{r}_b{^a}$ are almost complex structures obeying
\begin{subequations} \label{icplx}
\begin{gather} \label{iac}
\cs{r}_a{^c} \cs{s}_c{^b} + \cs{s}_a{^c} \cs{r}_c{^b} = -2\delta^{rs}
\delta_a^b,
\intertext{with vanishing Nijenhuis
concomitants,}
\label{nonij}
N(r,s)_{ab}{^c} \equiv \left\{ 2 \cs{r}_{[a}{^d} \p_{|d|} \cs{s}_{b]}{^c} - 2
\cs{r}_d{^c} \p_{[a} \cs{s}_{b]}{^d} \right\} + (r\leftrightarrow s) = 0.
\end{gather}
\end{subequations}
\end{ex}

From exercise~\ref{extsusy}, we learn that supersymmetry requires a complex
target space, with three anticommuting complex structures.
${\mathcal{N}}=4B$ supersymmetry is defined to have
a quaternionic target space,
\begin{equation} \label{quat}
\cs{r}_a{^c} \cs{s}_c{^b} = -\delta^{rs} \delta_a^b + \epsilon^{rst}
\cs{t}_a{^b}.
\end{equation}
This provides a natural $SU(2)$ structure, which will give rise to
self-dual rotations in the black hole context. Given such a
manifold, it is convenient to define the three exterior
derivatives $d^r$ by
\begin{equation} \label{dr} 
\begin{split}
d^r \omega &= (-1)^{p+1} \cs{r}d\cs{r}\omega
; \cs{r}\omega \equiv \frac{(-1)^p}{p!}
\cs{r}_{a_1}{^{b_1}}\cdots\cs{r}_{a_p}{^{b_p}}\omega_{b_1\dots b_p}
dX^{a_1}\wedge\dots\wedge dX^{a_p},
\\
\skipthis{&= \text{\small $\frac{(-1)^p}{p!}
\cs{r}_{a_1}{^{b_1}}\dots
\cs{r}_{a_{p+1}}{^{b_{p+1}}} \p_{b_1} \bigl( \cs{r}_{b_2}{^{c_1}}\dots
\cs{r}_{b_{p+1}}{^{c_p}} \omega_{c_1\dots
c_p}\bigr)dX^{a_1}\wedge\dots\wedge dX^{a_{p+1}}$} \\}
&= \frac{1}{p!} \left[ \cs{r}_a{^b} \p_b \omega_{c_1\dots c_p} -
p (\p_a \cs{r}_{c_1}{^b}) \omega_{bc_2\dots c_p}\right] dX^a \wedge
dX^{c_1} \wedge \dots \wedge dX^{c_p}
\end{split}
\end{equation}
where $\omega=\frac{1}{p!}\omega_{a_1\dots a_p}
dX^{a_1}\wedge\dots\wedge dX^{a_p}$ is a $p$-form.

\begin{ex} \label{exdr}
Show that, in complex coordinates adapted to $\cs{r}_a{^b}$, \hbox{$d^r
= i(\p-\Bar{\p})$}.
\end{ex}

Having defined the supersymmetry transformations using the quaternionic
structure of the manifold, we should now check that the action is
invariant.  We will simply quote the result~\cite{jmas1}.  The action is
invariant provided that%
\footnote{For the more general case, with a Clifford, but not a
quaternionic, structure see~\cite{cp,gps,hull}.}
\begin{subequations} \label{n=4susy}
\begin{gather}
\label{newnij}
\anti{d^r}{d^s}=0, \\
\label{newquat}
\cs{r}_a{^c} \cs{s}_c{^b} = -\delta^{rs} \delta_a^b + \epsilon^{rst}
\cs{t}_a{^b}, \\ \label{hermit}
g_{ab} = \cs{r}_a{^c} g_{cd} \cs{r}_b{^d} \, ( \forall \, r) \\
\label{dJ}
c = \frac{1}{6} c_{abc} dX^a \wedge dX^b \wedge dX^c = \half d^r \kf{r}
   \, ( \forall \, r); \kf{r} \equiv \half \cs{r}_a{^c} g_{cb}\, dX^a
   \wedge dX^b.
\end{gather}
\end{subequations}
Equation~\eqref{newnij} is secretly a restatement of
equation~\eqref{nonij} and
equation~\eqref{newquat} was our demand~\eqref{quat}.
The new
conditions are equations~\eqref{hermit} and~\eqref{dJ}.
Equation~\eqref{hermit} states that the metric is Hermitian with
respect to each complex structure.  Equation~\eqref{dJ} is a
highly nontrivial differential constraint between the complex
structures, which generalizes the hyperk\"{a}hler condition, and
from which the torsion $c_{abc}$ is uniquely determined.  It is
equivalent to the condition that the quaternionic structure be
covariantly constant:
\begin{equation} \label{dpi}
\nabla^+_a \cs{r}_b{^c} \equiv \nabla_a \cs{r}_b{^c} + c^c{_{ad}}
\cs{r}_b{^d}
- c^d{_{ab}} \cs{r}_d{^c} = 0.
\end{equation}
A manifold that satisfies the
conditions~\eqref{n=4susy} is known as a hyperk\"{a}hler with torsion
(HKT) (or sometimes weak HKT) manifold.%

\subsubsection{${\mathcal{N}}=4$B Superconformal Quantum Mechanics}
We must now find the further restrictions to a $D(2,1;\alpha)$-invariant
superconformal quantum mechanics.  Clearly, this will include
equations~\eqref{ldr}, \eqref{ius} and~\eqref{cfor1}.  The additional
restrictions are obtained by demanding the proper behaviour of the
R-symmetries, and are most easily phrased by defining the vector fields
\begin{equation} \label{defdr}
D^{rb} \equiv D^a \cs{r}_a{^b}.
\end{equation}
$D(2,1;\alpha$)-invariance then forces these to be Killing vectors
\begin{equation} \label{lgdr}
\lie{D^r}{g_{ab}} = 0,
\end{equation}
which also obey the $SU(2)$ algebra for some normalization
\begin{equation} \label{liedrds}
\com{\lie{D^r}{}}{\lie{D^s}{}} = -\frac{2}{\alpha+1} \epsilon^{rst}
\lie{D^t}{}.
\end{equation}
Equation~\eqref{liedrds} gives a geometric definition of $\alpha$.  Note
that because the normalization of $D^{ra}$ is specified by
equations~\eqref{ldr} and~\eqref{defdr}, $\alpha$ is unambiguous.  In fact,
equation~\eqref{liedrds} is not a sufficiently strong condition for the
proper closure of the algebra; we must have
\begin{equation} \label{lieIrds}
\lie{D^r}{\cs{s}_a{^b}} = -\frac{2}{\alpha+1} \epsilon^{rst}
\cs{t}_a{^b}.
\end{equation}
Equation~\eqref{lieIrds} implies equation~\eqref{liedrds}.

These are the necessary and sufficient conditions for the quantum mechanics
defined by~\eqref{l} to be $D(2,1;\alpha)$ superconformal.  They imply that
\begin{equation} \label{hktpot}
\alpha J^r = (\alpha+1)(d^r dK - \half \epsilon^{rst} d^s d^t K);
\end{equation}
{\em i.e.} that (at least for $\alpha\neq0$) the HKT metric is
described by a potential which is proportional to $K$. Actually,
as discussed in more detail in~\cite{jmas1}, when the three complex
structures are {\em simultaneously}
integrable, there is always a potential, but a
general HKT manifold admits a potential only under the conditions
given in~\cite{hktpot}.

A general (but not most general!) set of models can be obtained from a
function $L(X)$, where $X^a$ are coordinates on ${\mathbb{R}}^{4n}$, and the
$\cs{r}_a{^b}$ are given by the self-dual complex structures on
${\mathbb{R}}^4$ tensored with the $n$-dimensional identity matrix.%
\footnote{This implies that
the three complex structures are simultaneously integrable.  Hellerman and
Polchinski~\cite{hp} have recently shown how to relax this limitation by
generalizing the ${\mathcal{N}}=2$ superfield constraints of~\cite{ghr,dps}.}
If
(but not iff~--- in particular, this is not true of the system described in
section~\ref{bhscqm})
$L(X)$ also obeys
\begin{align} \label{lsr}
X^a \p_a L(X) &= h L(X), & X^a \cs{r}_a{^b} \p_b L(X) &= 0,
\end{align}
then we obtain a $D(2,1;\alpha=-\frac{h+2}{2})$-invariant model, with
\begin{subequations} \label{exhkt}
\begin{gather} \label{exg}
g_{ab} = \left(\delta^{c}_a \delta^d_b + \cs{r}_a{^c} \cs{r}_b{^d}\right)
\p_c \p_d L(X), \\
\label{exd}
D^a = \frac{2}{h} X^a, \\
\label{exk}
K = \frac{h+2}{2h} L,
\end{gather}
\end{subequations}
and $c_{abc}$ given by equation~\eqref{dJ}.  In an ${\mathcal N}=2$
superspace formalism, $X^a$ is a superfield obeying certain
constraints~\cite{ghr,dps} and the potential $L$ is the superspace
integrand~\cite{jmas1,hp,hull}.

\section{The Quantum Mechanics of a Test Particle in a
Reissner-Nordstr\"{o}m Background} \label{tp}

Our goal is to apply the results on superconformal quantum
mechanics to the quantum mechanics of a collection of
supersymmetric  black holes. As a warm-up in this section we
consider the problem of a quantum test particle moving in the
black hole geometry. The four-dimensional case was treated in
\cite{bhaf}, which will be followed and adapted to five dimensions
in this section.

Consider a five-dimensional extremal Reissner-Nordstr\"om black
hole of charge $Q$. The geometry of such a black hole is described
by the metric
\begin{subequations} \label{rn}
\begin{equation} \label{rng}
ds^2 = - \frac{dt^2}{\psi^2} + \psi d\vec{x}^2,
\end{equation}
and the gauge field
\begin{equation} \label{rna}
A = \psi^{-1} dt,
\end{equation}
\end{subequations}
where $\vec{x}$ is the ${\mathbb{R}}^4$ coordinate, and $\psi = 1 +
\frac{Q}{|\vec{x}^2|}$.  We have set $M_p = L_p = 1.$  The horizon in these
coordinates is at $|\vec{x}|=0$.

Introduce a test
particle with mass $m$ and charge $q$.  The particle action is
\begin{equation}\label{testaction}
S = -m \int d\tau + q \int A.
\end{equation}

Parametrize the particle's trajectory as $\vec{x} = \vec{x}(t).$  
Eventually we will require the test particle to be supersymmetric (by
imposing $q=m$).  A supersymmetric test particle at rest at a fixed
distance from the black hole, remains at rest,
so it is sensible to consider a test particle that moves slowly.
Accordingly we shall assume $|\Dot{\vec{x}}| \ll 1.$  
In this parametrization, we can make the following substitution:
\begin{equation}
d\vec{x} = \Dot{\vec{x}}dt,
\end{equation}
which allows us to rewrite \eqref{rn} to obtain the metric
\begin{equation}
ds^2 = - \frac{dt^2}{\psi^2} + \psi |\Dot{\vec{x}}|^2 dt^2.
\end{equation}
Now we solve the equation $ds^2 = -d\tau^2$ and find
\begin{equation}
d\tau = \frac{dt}{\psi} - \frac{1}{2}\psi^2 |\Dot{\vec{x}}|^2 dt +
{\cal{O}}(\Dot{x}^4),
\end{equation}
which is substituted into \eqref{testaction} to obtain the action,
\begin{equation}
S = -m \int (\frac{dt}{\psi} - \frac{1}{2}\psi^2 |\Dot{\vec{x}}|^2 dt) + q
\int \frac{dt}{\psi}.
\end{equation}

For a supersymmetric test particle, $m=q$, this action reduces to
\begin{equation}
S = \frac{m}{2} \int \psi^2 |\Dot{\vec{x}}|^2 dt.
\end{equation}

If the particle is near the horizon, at distances $r \ll \sqrt{Q},$ then we can approximate $\psi =
\frac{Q}{|\vec{x}|^2},$ so that

\begin{equation}
S_p = \frac{mQ^2}{2} \int dt \frac{|\Dot{\vec{x}}|^2}{|\vec{x}|^4},
\end{equation}
or, if we define a new quantity $\vec{y} = \frac{\vec{x}}{|\vec{x}|^2},$
then we see that we are actually in flat space:
\begin{equation}\label{dfgl}
S_p = \frac{mQ^2}{2} \int dt |\Dot{\vec{y}}|^2.
\end{equation}

Far from the black hole, spacetime and the moduli space look flat
once again.  Thus the moduli space can be described as two
asymptotically flat regions connected by a wormhole whose radius
scales as $\sqrt{Q}$. At low energies (relative to $M_p
/\sqrt{Q}$) the wavefunctions spread out and do not fit into the
wormhole. Hence the quantum mechanics is described by near and far
superselection sectors that decouple completely at low energies.

This geometry leads to a problem. Consider the near horizon
quantum theory. Given any fixed energy level $E,$ there are
infinitely many states of energy less than $E$. This suggests that
there are infinitely many states of a test particle localized near
the horizon of a black hole, which appears problematic for black
hole thermodynamics. The possibility of such states arises from
the large redshift factors near the horizon of a black hole.
Similar problems have been encountered in studies of ordinary
quantum fields in a black hole geometry.

The new observation of \cite{bhaf} is that this problem is in fact
equivalent to the problem encountered by DFF \cite{dff} in their
analysis of conformal quantum mechanics.  To see this equivalence
let $\rho$ denote the radial coordinate $|\vec{y}|$. The
Hamiltonian corresponding to \eqref{dfgl} is
\begin{equation}
H = \frac{1}{2mQ^2}(p_\rho^2 + \frac{4}{\rho^2} J^2).
\end{equation}
This is the DFF Hamiltonian of~\eqref{dffh} with $g = \frac{4}{mQ^2}J^2.$  The coordinate
$\rho$ grows infinite at the horizon.  Thus this potential pushes a particle
to the horizon whenever $J^2$ is nonzero.  Our problem of infinitely many states at low energies
is just the problem discussed by DFF.

Applying the DFF trick, as discussed in
section~\ref{dff}, provides the
solution to this problem.  We work in terms of $H+K$ rather than $H$, since
the former has a discrete spectrum of normalizable eigenstates.
There is an \slr\ symmetry generated by $H,D$ and $K$, where $D$ and $K$
are defined to be
\begin{subequations}
\begin{align}
D &= \half (\rho p_\rho + p_\rho \rho); \\
K &= \tfrac{1}{2}mQ^2\rho^2.
\end{align}
\end{subequations}
These generators satisfy equations \eqref{sl2r}.

The appearance of the \slr\ symmetry was not an accident. It arises
from the geometry of our spacetime.  Near the horizon, we find
that
\begin{equation}
ds^2 \to -\frac{r^4}{Q^2} dt^2 + \frac{Q}{r^2}dr^2 + Q d\Omega^2_3.
\end{equation}

We recognize this metric as that of $AdS_2 \times S^3.$
Introduce new coordinates $t^\pm = t \pm \frac{Q}{4r^2}$ on $AdS_2.$  Now
the metric can be written in the form
\begin{equation}\label{tcoords}
ds^2_2 = - \frac{Qdt^+dt^-}{(t^+ - t^-)^2}.
\end{equation}

The \slr\ isometry generators are then
\begin{subequations}
\begin{align}
h &= \frac{\p}{\p t^+} + \frac{\p}{\p t^-}, \\
d &= t^+ \frac{\p}{\p t^+} + t^- \frac{\p}{\p t^-}, \\
k &= (t^+)^2 \frac{\p}{\p t^+} + (t^-)^2 \frac{\p}{\p t^-}.
\end{align}
\end{subequations}
Here $h$ shifts the time coordinate, and $d$ rescales
all coordinates.

The \slr\ symmetry of the near-horizon particle action reflects the \slr\
isometry group of the near-horizon $AdS_2$ geometry.
As pointed out in \cite{bhaf,kall}, the trick of DFF to replace $H$ by $H+K$ has a nice interpretation in
$AdS_2.$  To understand it, we must first review $AdS_2$ geometry.

\subsection*{Interlude: $AdS_2$ Geometry}

On $AdS_2$, introduce global coordinates
$u^\pm$ defined in terms of the coordinates of \eqref{tcoords} by the
relation
\begin{equation}\label{adscoords}
t^\pm = \tan u^\pm.
\end{equation}
Then the $AdS_2$ metric takes the form
\begin{equation}
ds^2=-\frac{Q}{4}\frac{du^+ du^-}{\sin^2(u^+ - u^-)}.
\end{equation}

In these coordinates, the global time generator is
\begin{equation}
h+k=\frac{\p}{\p u^+} + \frac{\p}{\p u^-}.
\end{equation}

\begin{figure}[tb]
\begin{center}
\includegraphics[height=2.75in]{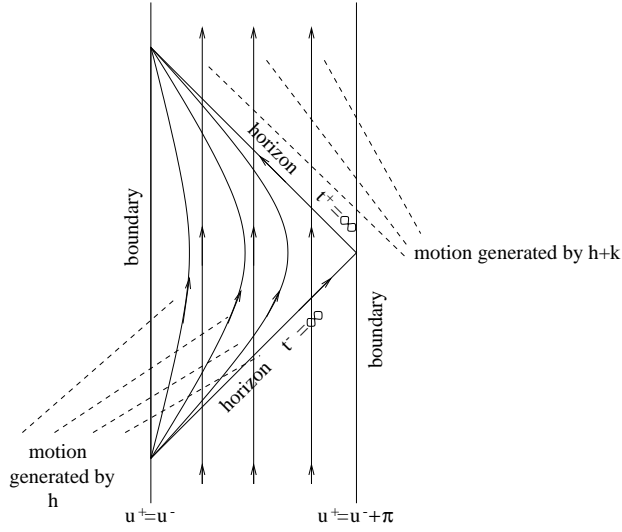}
\end{center}
\caption{The geometry of $AdS_2.$  The time conjugate to $h+k$ is a good
global coordinate.} \label{fig:hvsl0}
\end{figure}

In figure~\ref{fig:hvsl0} it is seen that the time coordinate conjugate to
$h$ is not a good global time coordinate on $AdS_2$, but the time
coordinate conjugate to $h+k$ is.  In fact, the generators $h$ and $d$
preserve the horizon, while $h+k$ preserves the boundary $u^+ =
u^- + \pi$ (the right boundary in figure~\ref{fig:hvsl0}).

So in conclusion the DFF trick has a beautiful geometric
interpretation in the black hole context. It is simply a
coordinate transformation to ``good'' coordinates on $AdS_2.$

\section{Quantum Mechanics on the Black Hole Moduli Space} \label{bhms}

\subsection{The black hole moduli space metric}

In this section we will consider  five-dimensional ${\cal N}=1$
supergravity with a single $U(1)$ charge coupled to the
graviphoton and
no vector multiplets.%
\footnote{ Adding neutral hypermultiplets would not affect the
discussion, since they decouple. Since these lectures were given,
the case with additional vector multiplets was solved
in~\cite{gp}, and the four-dimensional case was solved in
\cite{alexmark}. The supersymmetry of cases with more than eight initial
supersymmetries~\cite{dkjm,jmfour} has not been worked out. } We will use
units with
$M_p=L_p=1$. The action is
\begin{equation}
\label{puresugra}
S =  \int d^5x \sqrt{g} \bigl[R - \frac{3}{4} F^2]
 + \frac{1}{2} \int A \wedge F \wedge F +\text{fermions}.
\end{equation}
We can also get this from M-theory compactified on a Calabi-Yau
with $b_2=1$ (the simplest example of such a threefold is the
quintic). The black holes are then M2-branes wrapping Calabi-Yau
two-cycles.

This system has a solution
describing  $N$ static extremal black holes
\begin{subequations} \label{totbhsoln}
\begin{gather}
ds^2 = -\psi^{-2} dt^2 + \psi d\vec{x}^2,
\label{bhsoln} \\
A = \psi^{-1} dt,
\label{bhsolnA}
\end{gather}
({\em cf.} equation~\eqref{rn}) where $\psi$ is the harmonic function on ${\mathbb{R}}^4$
\begin{equation}
\label{bhsolnpsi}
\psi = 1 + \sum_{A=1}^N \frac{Q_A}{|\vec{x}-\vec{x}_A|^2},
\end{equation}
\end{subequations}
and $\vec{x}_A$ is the ${\mathbb{R}}^4$ coordinate of the
$A^{\text{\underline{th}}}$ black hole, whose charge is $Q_A$.
Another picture of these holes is  M2-branes
wrapping Calabi-Yau cycles.
The space of solutions is called the moduli space,
which is parametrized by the $4N$  collective coordinates $\vec{x}_A$.
The slow motion of such black holes
is governed by the moduli space metric $G_{AB},$
so that the low energy effective
action takes the form
\begin{equation}
S=\frac{1}{2}\int dt \Dot x^A \Dot x^B G_{AB}.
\label{modspmet}
\end{equation}
Note that due to the no-force condition there is no
potential term in the action, and since
$|\Dot{\vec{x}}_A| \ll 1,$ the higher order corrections can be neglected.

The first  calculation of the moduli space metric
of the four-dimensional Reissner-Nordstr\"{o}m  black holes
was performed in \cite{fe,gr} and was generalized to
dilaton black holes in \cite{shir}.
The metric on the moduli space
for the five-dimensional black holes (\ref{totbhsoln})
was derived
in \cite{jmas2}.
In order to find this metric, one
starts with the following
ansatz describing the linear order perturbation
of the black hole
solution~\eqref{totbhsoln}
\begin{subequations} \label{totpertbh}
\begin{gather}
\label{pertbh}
ds^2 = -\psi^{-2} dt^2 + \psi d\vec{x}^2 + 2 \psi^{-2} \vec{R}
\cdot d\vec{x} dt, \\
\label{pertbhA}
A = \psi^{-1} dt + (\vec{P} - \psi^{-1} \vec{R}) \cdot d\vec{x},
\end{gather}
\end{subequations}
where $\vec{P}$ and $\vec{R}$ are quantities that are first order in
velocities.
In equation~\eqref{bhsolnpsi}, $\vec{x}_A$ is replaced with
$\vec{x}_A+\vec{v}_At$.
This is the most general Galilean-invariant ansatz to linear
order.
Then (roughly) one uses the equations of motion to solve for
$\vec{P}$ and $\vec{R}$.
Inserting this into the five-dimensional supergravity action
gives the following result \cite{jmas2} for the action:
\begin{equation}
S=\frac{1}{2}\int dt \Dot x^A \Dot x^B G_{AB}=
\frac{1}{4} \int dt \Dot x^{Ak} \Dot x^{Bl}
(\delta ^i_k \delta^j_l + I^{ri}_k I^{rj}_l)
\partial_{Ai} \partial_{Bj} L,
\label{act}
\end{equation}
where
\begin{equation}
\label{ldef}
L=-\int d^4x \psi^3,
\end{equation}
with
\begin{equation}
\label{pgl}
\psi = 1+\sum_{A=1}^N \frac{Q_A}{|\vec{x}-\vec{x}_A|^2},
\end{equation}
and $I^r$ is a triplet of self-dual complex structures on
${\mathbb{R}}^4$ obeying equation~\eqref{quat}.
This Lagrangian has ${\cal N}=4$ supersymmetry when
Hermitian fermions $\lambda ^{Ai}=\lambda^{Ai\dagger}$ are added.

\begin{figure}[tb]
\begin{center}
\includegraphics[height=2in]{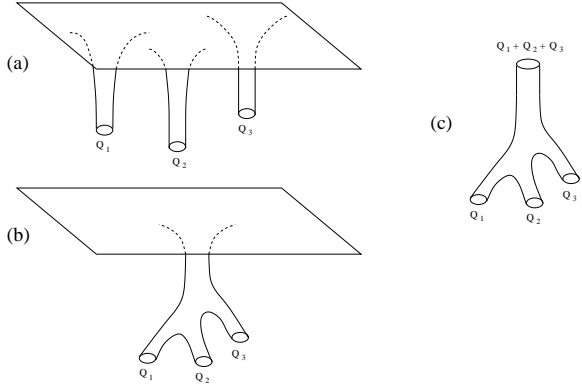}
\end{center}
\caption{ (a) Widely separated black holes.
(b) Near-coincident black holes.
(c) The near-horizon limit.}
\label{fig:nhst}
\end{figure}

\subsection{The Near-Horizon Limit}

\subsubsection{Spacetime geometry}

Taking the near-horizon limit of \eqref{bhsoln}
corresponds to neglecting the constant term in \eqref{bhsolnpsi}.
In figure~\ref{fig:nhst} we have illustrated the resultant
spatial geometry at a moment of fixed time for three black holes.
Before the limit is taken (figure~\ref{fig:nhst}a), the geometry has an
asymptotically flat region at large $|\vec{x}|$.  Near the limit
(figure~\ref{fig:nhst}b), as the origin
is approached along a spatial trajectory,
a single ``throat''
approximating that of a charge $\sum Q_A$
black hole is encountered. This throat region is
an $AdS_2\times S^3$ geometry with radii of order $\sqrt {\sum Q_A}$.
As one moves deeper inside the throat towards the horizon, the throat
branches into smaller throats, each of which has smaller charge and
correspondingly smaller radii.
Eventually there are $N$ branches with
charge $Q_A$. At the end of each of these branches is an event horizon.
When the limit is achieved (figure~\ref{fig:nhst}c),
the asymptotically flat region moves off to
infinity.
Only the charge $\sum Q_A$  ``trunk'' and the many branches remain.

\subsubsection{Moduli space geometry}

\begin{figure}[tb]
\begin{center}
\includegraphics[height=1.9in]{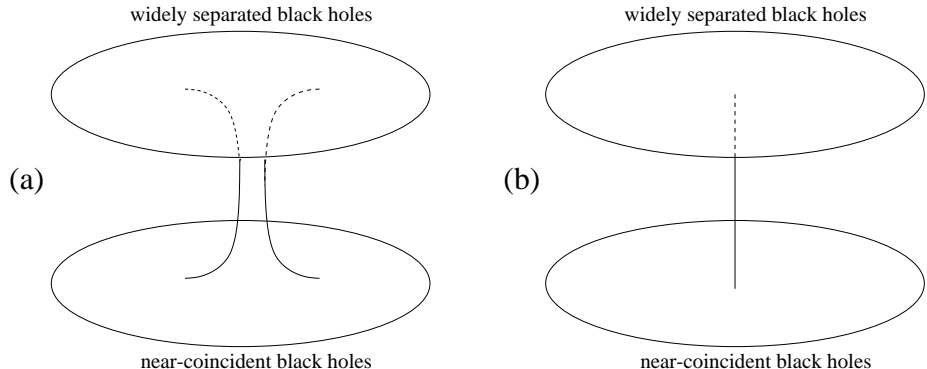}
\end{center}
\caption{
(a) Regions of the two-black hole moduli space.
(b) The near-horizon limit.}
\label{fig:nhms}
\end{figure}

It is also interesting to consider the near-horizon limit
of the moduli space geometry.
The metric is again given by \eqref{act},
where one should neglect the constant term in the harmonic function
\eqref{pgl}.
This is illustrated in
figure~\ref{fig:nhms} for the case of two black holes.
Near the limit there is an
asymptotically flat ${\mathbb{R}}^{4N}$ region corresponding to
all $N$ black holes being widely separated. This is connected
to the near-horizon
region where the black holes are strongly interacting, by
tubelike regions which become longer and thinner as the
limit is approached. When the limit is achieved, the
near-horizon region is severed from the tubes and the asymptotically
flat region.


\subsection{Conformal Symmetry} \label{bhscqm}

The near-horizon
quantum mechanics has an \slr\ conformal symmetry.
The
dilations $D$ and special conformal transformations $K$
are generated by
\begin{gather}
D=- \frac{1}{2}(x^{Ai} P_{Ai}+ h.c.), \\
K=6 \pi^2\sum_{A \neq B}^N
\frac{Q_A^2Q_B}{|\vec{x}_A-\vec{x}_B|^2}.
\end{gather}

By splitting the potential $L$
appearing in the metric~\eqref{act}
into pieces representing the 1-body, 2-body and 3-body interactions,
one can show \cite{jmas2}
that the conditions~\eqref{lgdr} and~\eqref{lieIrds}
are satisfied. Thus
the \slr\ symmetry can be extended to
the full $D(2,1;0)$ superconformal symmetry
as was described in section~\ref{d21a}.
This group is
the special case of the $D(2,1;\alpha)$ superconformal groups
for which there is an $SU(1,1|2)$ subgroup
(in fact, $D(2,1;0) \cong  SU(1,1|2) \rtimes SU(2)$), 
in agreement with \cite{gmt}.

So we have seen that there are noncompact regions of the
near-horizon moduli space corresponding to
coincident black holes.
These regions are eliminated by the
potential $K$ in the modified Hamiltonian $L_0=\frac{1}{2}(H+K)$,
which is singular at the boundary of the noncompact regions.
$L_0$ has a well defined spectrum with discrete eigenstates.
A detailed description of the quantum states of this system
remains to be found \cite{rsn}.

\section{Discussion} \label{conc}

Let us recapitulate. We have found that at low energies the
quantum mechanics of $N$ black holes divides into superselection sectors.
One sector describes the dynamics of widely separated, non-interacting black
holes.
The other ``near horizon'' sector describes highly redshifted,
near-coincident black holes and has an enhanced superconformal
symmetry. Since they completely
decouple from widely separated black holes,  states of the near horizon theory
are multi-black hole bound states.

It is instructive to compare this to an M-theoretic description of
these black holes. In Calabi-Yau compactification of M theory to
five dimensions, the black holes are described by M2-branes
multiply wrapped around holomorphic cycles of the Calabi-Yau. In
principle all the black hole  microstates are described by quantum
mechanics on the M2-brane moduli space, which at low energies
should be the dual CFT$_1$ living on the boundary of $AdS_2$
\cite{jm}. In practice so far this problem has not been tractable.
This moduli space has what could be called (in a slight abuse of
terminology) a Higgs branch and a Coulomb branch. This Higgs
branch is a sigma model whose target is the moduli space of a
single multiply wrapped M2-brane worldvolume in the
Calabi-Yau. In the Coulomb branch the M2-brane has fragmented into
multiple pieces, and the branch is parametrized by the M2-brane
locations. At finite energy the Coulomb branch connects to the
Higgs branch at singular points where the M2-brane worldvolume
degenerates.

At first one might think that the considerations of this paper
correspond to the Coulomb branch, since the multi-black hole
moduli space is parametrized by the black hole locations. However
it is not so simple. The fact that the near horizon sector
decouples from the sector describing non-interacting black holes
strongly suggests that it is joined to the Higgs branch. Indeed in
the D1/D5 black hole, there is a similar near-horizon region of
the Coulomb branch which is not only joined to but is in fact a
dual description of the singular regions of the Higgs branch
\cite{ofone,mms,oftwo,bv,ab}. We conjecture there is a similar story
here: the near-horizon, multi-black hole quantum mechanics is dual
to at least part of the Higgs branch of multiply wrapped
M2-branes. Near-horizon microstates should therefore account for
at least some of the internal black hole microstates. Exactly how
much of the black hole microstructure is accounted for in this way
remains to be understood.

\section*{Acknowledgments}

We thank F. Cachazo, R. Jackiw, J. Maldacena, A. Maloney, G. Papadopoulos,
L. Thorlacius, P. Townsend 
and especially M. Spradlin
for enlightening discussions and communication.  The research reviewed here
was supported in part by NSERC and DOE grant DE-FGO2-91ER40654.  We thank
the organizers and especially L\'{a}rus Thorlacius for an excellent school,
the
participants for their stimulating questions, and NATO for financial
support at the school.

\appendix

\section{Differential Geometry with Torsion} \label{torsion}

In this appendix, we give a brief summary of differential calculus with
torsion, for the reader who is frustrated by the usual absence of such a
discussion in most general relativity books.%
\footnote{One excellent reference for physicists is~\cite{nn}.}
Recall~\cite{wald} that the covariant derivative of a tensor is given in
terms of the (not necessarily symmetric) connection $C^c_{ab}$.  The
torsion $c^c{_{ab}}$ is just the antisymmetric
part of the connection:
\begin{equation} \label{deft}
c^c{_{ab}} \equiv C^c_{[ab]} = \half (C^c_{ab}-C^c_{ba}).
\end{equation}
Either by direct computation, or by recalling that the difference between
two connections is a tensor, one finds that the torsion is a
true tensor.  Of course, the torsion does contribute to the curvature
tensor, and we remind the reader that many of the familiar symmetries of
the curvature tensor are not obeyed in the presence of torsion.
\skipthis{
For example, the differential Bianchi identity is
\begin{equation} \label{dbianchi}
\partial_{[a} R_{bc]de} = ...
\end{equation}
}
Also, if the symmetric part of the connection is given by the
Levi-Civita connection, then the full connection annihilates the metric iff
the fully covariant torsion tensor $c_{abc} = g_{ad} c^d{_{bc}}$ is
completely antisymmetric.

Hopefully, the preceding paragraph was familiar.  We now discuss the
torsion in a tangent space formalism.  As usual, the first step is to
define the vielbein $e_a^\alpha$, which is a basis of
cotangent space vectors, labelled by $\alpha=1,\dots,n$, where $n$ is the
dimension of the manifold, obeying
\begin{equation} \label{defe}
\delta_{\alpha\beta} e^\alpha_a e^{\beta}_b = g_{ab}.
\end{equation}
The vielbein $e_a^\alpha$, and the inverse vielbein $e_\alpha^a$ which
obeys
\begin{equation} \label{defie}
e_\alpha^a e_a^\beta = \delta_\alpha^\beta,
\end{equation}
can then be used to map tensors into the tangent space; {\em e.g.} $V^\alpha
\equiv V^a e_a^\alpha$.

The connection one-form $\Omega_a{^\alpha}{_\beta}$
is
defined by demanding that the vielbein is covariantly constant:
\begin{equation} \label{de=0}
\nabla_a e_b^\alpha \equiv \p_a e_b^\alpha + \Omega_a{^\alpha}{_\beta}
e_b^\beta - C_{ab}^c e_c^\alpha = 0.
\end{equation}
Note that equation~\eqref{de=0} is valid for any choice of connection, and
does not imply that the metric is covariantly constant.  The metric is
covariantly constant iff $\delta_{\alpha\beta}$ is covariantly constant,
which in
turn holds iff the connection one-form $\Omega_{a\alpha\beta}$ is
antisymmetric in
the tangent space indices, where we have lowered the middle index using the
tangent space metric $\delta_{\alpha\beta}$.  In other words, the
familiar
antisymmetry of the connection one-form~\cite{wald} exists if and only if the metric is
covariantly
constant, whether or not there is torsion.

Equation~\eqref{de=0} is easily solved for the connection one-form,
giving
\begin{equation} \label{defo}
\Omega_a{^\alpha}{_\beta} = e_b^\alpha \p_a e^b_\beta + C_{ab}^c e_c^\alpha
e^b_\beta.
\end{equation}
An immediate corollary of this, and the fact that the difference of two
connections $C^c_{ab}$ and ${C'}^c_{ab}$ is a tensor, is that the
difference between two connection one-forms is a tensor, and is, in
fact, the same tensor as $C^c_{ab}-{C'}^c_{ab}$, but with the $b$ and $c$
indices lifted to the tangent bundle.%
\footnote{In this discussion, we are assuming that a vielbein has been
chosen once and for all; we do not consider the effect of changing frames
or coordinates.}

The unique torsion-free connection one-form which annihilates
the metric ({\em i.e.} that obtained from equation~\eqref{de=0} using the
Levi-Civita connection) is known as the spin connection, and is usually
denoted $\omega_a{^\alpha}{_\beta}$.  Given a completely antisymmetric
torsion $c_{abc}=c_{[abc]}$, as in the first paragraph of this appendix, we
define the connection one-form
\begin{equation} \label{O+}
\Omega^+_a{^\alpha}{_\beta} = \omega_a{^\alpha}{_\beta} +
c{^\alpha}{_{a\beta}},
\end{equation}
where, of course, any required mapping between the tangent bundle and the
spacetime is achieved by contracting with the vielbein.

As usual, spinors $\psi$ are defined on the tangent bundle, and their
covariant
derivative is given by
\begin{equation} \label{dpsi}
\nabla_a \psi = \p_a \psi - \frac{1}{4} \Omega_{a\alpha\beta}
\gamma^{\alpha \beta} \psi,
\end{equation}
where $\gamma^{\alpha\beta}\equiv
\half\com{\gamma^\alpha}{\gamma^\beta}$ is a commutator of $SO(n)$
$\gamma$-matrices, which satisfy
\hbox{$\anti{\gamma^\alpha}{\gamma^\beta}=2\delta^{\alpha\beta}$}.

\end{document}